\documentclass[twoside,reqno]{bjp}
\usepackage{epsfig,cite}
\usepackage{graphics}
\usepackage{amssymb,amsmath}
\usepackage{times}
\begin{document}

\sloppy \raggedbottom
\setcounter{page}{1}

%
%
%
%
\newcommand\rf[1]{(\ref{eq:#1})}
\newcommand\lab[1]{\label{eq:#1}}
\newcommand\nonu{\nonumber}
\newcommand\br{\begin{eqnarray}}
\newcommand\er{\end{eqnarray}}
\newcommand\be{\begin{equation}}
\newcommand\ee{\end{equation}}
\newcommand\eq{\!\!\!\! &=& \!\!\!\! }
\newcommand\foot[1]{\footnotemark\footnotetext{#1}}
\newcommand\lb{\lbrack}
\newcommand\rb{\rbrack}
\newcommand\llangle{\left\langle}
\newcommand\rrangle{\right\rangle}
\newcommand\blangle{\Bigl\langle}
\newcommand\brangle{\Bigr\rangle}
\newcommand\llb{\left\lbrack}
\newcommand\rrb{\right\rbrack}
\newcommand\Blb{\Bigl\lbrack}
\newcommand\Brb{\Bigr\rbrack}
\newcommand\lcurl{\left\{}
\newcommand\rcurl{\right\}}
\renewcommand\({\left(}
\renewcommand\){\right)}
\renewcommand\v{\vert}                     
\newcommand\bv{\bigm\vert}               
\newcommand\Bgv{\;\Bigg\vert}            
\newcommand\bgv{\bigg\vert}              
\newcommand\lskip{\vskip\baselineskip\vskip-\parskip\noindent}
\newcommand\mskp{\par\vskip 0.3cm \par\noindent}
\newcommand\sskp{\par\vskip 0.15cm \par\noindent}
\newcommand\bc{\begin{center}}
\newcommand\ec{\end{center}}
\newcommand\Lbf[1]{{\Large \textbf{{#1}}}}
\newcommand\lbf[1]{{\large \textbf{{#1}}}}




\newcommand\tr{\mathop{\mathrm tr}}                  
\newcommand\Tr{\mathop{\mathrm Tr}}                  
\newcommand\partder[2]{\frac{{\partial {#1}}}{{\partial {#2}}}}
\newcommand\partderd[2]{{{\partial^2 {#1}}\over{{\partial {#2}}^2}}}
\newcommand\partderh[3]{{{\partial^{#3} {#1}}\over{{\partial {#2}}^{#3}}}}
\newcommand\partderm[3]{{{\partial^2 {#1}}\over{\partial {#2} \partial{#3} }}}
\newcommand\partderM[6]{{{\partial^{#2} {#1}}\over{{\partial {#3}}^{#4}{\partial {#5}}^{#6} }}}          
\newcommand\funcder[2]{{{\delta {#1}}\over{\delta {#2}}}}
\newcommand\Bil[2]{\Bigl\langle {#1} \Bigg\vert {#2} \Bigr\rangle}  
\newcommand\bil[2]{\left\langle {#1} \bigg\vert {#2} \right\rangle} 
\newcommand\me[2]{\left\langle {#1}\right|\left. {#2} \right\rangle} 

\newcommand\sbr[2]{\left\lbrack\,{#1}\, ,\,{#2}\,\right\rbrack} 
\newcommand\Sbr[2]{\Bigl\lbrack\,{#1}\, ,\,{#2}\,\Bigr\rbrack}
\newcommand\Gbr[2]{\Bigl\lbrack\,{#1}\, ,\,{#2}\,\Bigr\} }
\newcommand\pbr[2]{\{\,{#1}\, ,\,{#2}\,\}}       
\newcommand\Pbr[2]{\Bigl\{ \,{#1}\, ,\,{#2}\,\Bigr\}}  
\newcommand\pbbr[2]{\lcurl\,{#1}\, ,\,{#2}\,\rcurl}




\renewcommand\a{\alpha}
\renewcommand\b{\beta}
\renewcommand\c{\chi}
\renewcommand\d{\delta}
\newcommand\D{\Delta}
\newcommand\eps{\epsilon}
\newcommand\vareps{\varepsilon}
\newcommand\g{\gamma}
\newcommand\G{\Gamma}
\newcommand\grad{\nabla}
\newcommand\h{\frac{1}{2}}
\renewcommand\k{\kappa}
\renewcommand\l{\lambda}
\renewcommand\L{\Lambda}
\newcommand\m{\mu}
\newcommand\n{\nu}
\newcommand\om{\omega}
\renewcommand\O{\Omega}
\newcommand\p{\phi}
\newcommand\vp{\varphi}
\renewcommand\P{\Phi}
\newcommand\pa{\partial}
\newcommand\tpa{{\tilde \partial}}
\newcommand\bpa{{\bar \partial}}
\newcommand\pr{\prime}
\newcommand\ra{\rightarrow}
\newcommand\lra{\longrightarrow}
\renewcommand\r{\rho}
\newcommand\s{\sigma}
\renewcommand\S{\Sigma}
\renewcommand\t{\tau}
\renewcommand\th{\theta}
\newcommand\bth{{\bar \theta}}
\newcommand\Th{\Theta}
\newcommand\z{\zeta}
\newcommand\ti{\tilde}
\newcommand\wti{\widetilde}
\newcommand\twomat[4]{\left(\begin{array}{cc}  
{#1} & {#2} \\ {#3} & {#4} \end{array} \right)}
\newcommand\threemat[9]{\left(\begin{array}{ccc}  
{#1} & {#2} & {#3}\\ {#4} & {#5} & {#6}\\
{#7} & {#8} & {#9} \end{array} \right)}


\newcommand\cA{{\mathcal A}}
\newcommand\cB{{\mathcal B}}
\newcommand\cC{{\mathcal C}}
\newcommand\cD{{\mathcal D}}
\newcommand\cE{{\mathcal E}}
\newcommand\cF{{\mathcal F}}
\newcommand\cG{{\mathcal G}}
\newcommand\cH{{\mathcal H}}
\newcommand\cI{{\mathcal I}}
\newcommand\cJ{{\mathcal J}}
\newcommand\cK{{\mathcal K}}
\newcommand\cL{{\mathcal L}}
\newcommand\cM{{\mathcal M}}
\newcommand\cN{{\mathcal N}}
\newcommand\cO{{\mathcal O}}
\newcommand\cP{{\mathcal P}}
\newcommand\cQ{{\mathcal Q}}
\newcommand\cR{{\mathcal R}}
\newcommand\cS{{\mathcal S}}
\newcommand\cT{{\mathcal T}}
\newcommand\cU{{\mathcal U}}
\newcommand\cV{{\mathcal V}}
\newcommand\cX{{\mathcal X}}
\newcommand\cW{{\mathcal W}}
\newcommand\cY{{\mathcal Y}}
\newcommand\cZ{{\mathcal Z}}

\newcommand{\nit}{\noindent}
\newcommand{\ct}[1]{\cite{#1}}
\newcommand{\bib}[1]{\bibitem{#1}}

\newcommand\PRL[3]{\textsl{Phys. Rev. Lett.} \textbf{#1} (#2) #3}
\newcommand\NPB[3]{\textsl{Nucl. Phys.} \textbf{B#1} (#2) #3}
\newcommand\NPBFS[4]{\textsl{Nucl. Phys.} \textbf{B#2} [FS#1] (#3) #4}
\newcommand\CMP[3]{\textsl{Commun. Math. Phys.} \textbf{#1} (#2) #3}
\newcommand\PRD[3]{\textsl{Phys. Rev.} \textbf{D#1} (#2) #3}
\newcommand\PLA[3]{\textsl{Phys. Lett.} \textbf{#1A} (#2) #3}
\newcommand\PLB[3]{\textsl{Phys. Lett.} \textbf{#1B} (#2) #3}
\newcommand\CQG[3]{\textsl{Class. Quantum Grav.} \textbf{#1} (#2) #3}
\newcommand\JMP[3]{\textsl{J. Math. Phys.} \textbf{#1} (#2) #3}
\newcommand\PTP[3]{\textsl{Prog. Theor. Phys.} \textbf{#1} (#2) #3}
\newcommand\SPTP[3]{\textsl{Suppl. Prog. Theor. Phys.} \textbf{#1} (#2) #3}
\newcommand\AoP[3]{\textsl{Ann. of Phys.} \textbf{#1} (#2) #3}
\newcommand\RMP[3]{\textsl{Rev. Mod. Phys.} \textbf{#1} (#2) #3}
\newcommand\PR[3]{\textsl{Phys. Reports} \textbf{#1} (#2) #3}
\newcommand\FAP[3]{\textsl{Funkt. Anal. Prilozheniya} \textbf{#1} (#2) #3}
\newcommand\FAaIA[3]{\textsl{Funct. Anal. Appl.} \textbf{#1} (#2) #3}
\newcommand\TAMS[3]{\textsl{Trans. Am. Math. Soc.} \textbf{#1} (#2) #3}
\newcommand\InvM[3]{\textsl{Invent. Math.} \textbf{#1} (#2) #3}
\newcommand\AdM[3]{\textsl{Advances in Math.} \textbf{#1} (#2) #3}
\newcommand\PNAS[3]{\textsl{Proc. Natl. Acad. Sci. USA} \textbf{#1} (#2) #3}
\newcommand\LMP[3]{\textsl{Letters in Math. Phys.} \textbf{#1} (#2) #3}
\newcommand\IJMPA[3]{\textsl{Int. J. Mod. Phys.} \textbf{A#1} (#2) #3}
\newcommand\IJMPD[3]{\textsl{Int. J. Mod. Phys.} \textbf{D#1} (#2) #3}
\newcommand\TMP[3]{\textsl{Theor. Math. Phys.} \textbf{#1} (#2) #3}
\newcommand\JPA[3]{\textsl{J. Physics} \textbf{A#1} (#2) #3}
\newcommand\JSM[3]{\textsl{J. Soviet Math.} \textbf{#1} (#2) #3}
\newcommand\MPLA[3]{\textsl{Mod. Phys. Lett.} \textbf{A#1} (#2) #3}
\newcommand\JETP[3]{\textsl{Sov. Phys. JETP} \textbf{#1} (#2) #3}
\newcommand\JETPL[3]{\textsl{ Sov. Phys. JETP Lett.} \textbf{#1} (#2) #3}
\newcommand\PHSA[3]{\textsl{Physica} \textbf{A#1} (#2) #3}
\newcommand\PHSD[3]{\textsl{Physica} \textbf{D#1} (#2) #3}
\newcommand\JPSJ[3]{\textsl{J. Phys. Soc. Jpn.} \textbf{#1} (#2) #3}
\newcommand\JGP[3]{\textsl{J. Geom. Phys.} \textbf{#1} (#2) #3}

\newcommand\Xdot{\stackrel{.}{X}}
\newcommand\xdot{\stackrel{.}{x}}
\newcommand\ydot{\stackrel{.}{y}}
\newcommand\yddot{\stackrel{..}{y}}
\newcommand\rdot{\stackrel{.}{r}}
\newcommand\rddot{\stackrel{..}{r}}
\newcommand\vpdot{\stackrel{.}{\varphi}}
\newcommand\vpddot{\stackrel{..}{\varphi}}
\newcommand\tdot{\stackrel{.}{t}}
\newcommand\zdot{\stackrel{.}{z}}
\newcommand\etadot{\stackrel{.}{\eta}}
\newcommand\udot{\stackrel{.}{u}}
\newcommand\vdot{\stackrel{.}{v}}
\newcommand\rhodot{\stackrel{.}{\rho}}
\newcommand\xdotdot{\stackrel{..}{x}}
\newcommand\ydotdot{\stackrel{..}{y}}



\title{Dynamical Volume Element in Scale-Invariant and Supergravity Theories
\thanks{Invited talk at Second Bulgarian National Congress in Physics, Sept. 2013}}

\begin{start}
\author{\underline{E.~Guendelman}}{1}, \coauthor{E.~Nissimov}{2},
\coauthor{S.~Pacheva}{2}, \coauthor{M.~Vasihoun}{1}

\address{Department of Physics, Ben-Gurion Univ. of the Negev,
Beer-Sheva 84105, Israel}{1}

\address{Institute of Nuclear Research and Nuclear Energy,
Bulg. Acad. Sci., Sofia 1784, Bulgaria}{2}

\runningheads{\underline{E.~Guendelman}, E.~Nissimov, S.~Pacheva, M.~Vasihoun}{Dynamical Volume 
Element in Scale-Invariant and Supergravity Theories}

\received{}

\begin{Abstract}
The use in the action integral of a volume element of the form $\Phi d^{D}x$, where
$\Phi$ is a metric-independent measure density, can yield new interesting results in all
types of known generally coordinate-invariant theories: (1) 4-D theories of 
gravity plus matter fields; (2) reparametrization invariant theories of extended
objects (strings and branes); (3) supergravity theories. In case (1) we obtain interesting 
insights concerning the cosmological constant problem, inflation and quintessence
without the fifth force problem. In case (2) the above formalism leads to dynamically 
induced tension and to string models of non-abelian confinement.
In case (3), we show that the modified-measure supergravity generates an arbitrary 
dynamically induced cosmological constant, \textsl{i.e.}, a new mechanism of dynamical 
supersymmetry breaking. 
\end{Abstract}


\end{start}

\section[]{Introduction}
In Refs.\ct{G6,G7} we have studied a new class of gravity theories based on the
idea that the action integral may contain a new metric-independent measure of
integration. For example, in $D=4$ space-time dimensions the new measure
density can be built out of four auxiliary scalar fields $\varphi^{i}\,$ ($i=1,2,3,4$):
\be
\Phi (\vp) =
\frac{1}{4!}\varepsilon^{\m\n\k\l}\varepsilon_{ijkl}\partial_{\m}\varphi^{i}
\partial_{\n}\varphi^{j}\partial_{\k}\varphi^{k} \partial_{\l}\varphi^{l} \; .
\lab{Phi}
\ee
$\Phi (\vp)$ is a scalar density under general coordinate
transformations. Here we will discuss three applications:

\begin{itemize}
\item
(i) Study of $D=4$-dimensional models of gravity and matter fields containing the new 
measure of integration \rf{Phi}, which appears to be promising candidates for resolution
of the dark energy and dark matter problems, the fifth force problem, {\em etc}.
\item
(ii) Study of a new type of string and brane models based on employing
of a modified world-sheet/world-volume integration measure. It allows for
the appearance of new types of objects and effects like, for example, a
spontaneously induced variable string tension.
\item
(iii) Studying modified supergravity models. Here we will find some outstanding new 
features: (a) the cosmological constant arises as an arbitrary integration constant, 
totally unrelated to the original parameters of the action, 
and (b) dynamical spontaneous breakdown of 
supersymmetry invariance. 
\end{itemize}

\section{Gravity and Cosmology Two Measures Theory}.
We consider action principle of the following general form:
\be
S = \int L_{1}\Phi d^{4}x +\int L_{2}\sqrt{-g}d^{4}x \; ,
\lab{S}
\ee
including two Lagrangians $ L_{1}$ and $L_{2}$ and two
measures of the volume elements ($\Phi d^{4}x$ and the standard one 
$\sqrt{-g}d^{4}x$, respectively). In constructing field theory with the action
\rf{S} we make only two basic additional  assumptions:

$\phantom{aa}$(A) $L_{1}$ and $L_{2}$ are independent of the measure fields
$\varphi_{i}$. Then the action \rf{S} is invariant under volume-preserving 
diffeomorphisms on the target space of the latter \ct{G6}. Besides, it is
invariant (up to an integral of a total divergence) under the
infinite-dimensional group of shifts of the measure fields
$\varphi^{i}$: $\varphi^{i}\rightarrow\varphi^{i}+f^{i}(L_{1})$,
where $f^{i}(L_{1})$ is an arbitrary differentiable function of
the Lagrangian density $L_{1}$.

$\phantom{aa}$(B) We proceed in the first-order formalism where all fields,
including the metric $g_{\m\n}$ (or the vierbeins ${e}^a_{\m}$),
connection coefficients (or spin-connection $\omega_{\m\,ab}$)
and the measure fields $\varphi^{i}$ are \textsl{a priori} independent dynamical
variables. All the relations between them follow subsequently as a result of
the equations of motion. 

The field theory based on the listed assumptions we call
``Two Measures Theory'' (TMT).
It turns out that the measure fields $\varphi_{i}$ affect the
theory only via the ratio of the two measure densities $\chi \equiv \Phi /\sqrt{-g}$,
which is a scalar field. It is determined by a constraint in
the form of an algebraic equation, which is precisely a consistency
condition of the equations of motion.  {\em This constraint determines
$\chi$ in terms of the fermion density and scalar fields}.

By an appropriate change of the dynamical variables, 
consisting of a conformal rescaling of the metric and a
multiplicative redefinitions of the fermion fields, one can
formulate the theory as a model in a Riemannian (or Riemann-Cartan)
space-time. The corresponding conformal frame we call ``the
Einstein frame''. 

We have started a detailed study of gravity-matter models
with a general form for $L_1$ and $L_2$  such that the action \rf{S} possesses  both 
non-Abelian gauge symmetry as well as scale symmetry. For brevity, in a schematic
form $L_1$ can be represented as ($\k^2=8\pi G_N$, $G_N$ -- Newton constant):
\be
L_1=e^{\alpha\phi /M_{p}}\Bigl\lb\frac{1}{\k^2}R(\om,e)
-\frac{1}{2}g^{\m\n}\phi_{,\mu}\phi_{,\nu}+({\rm Higgs})+({\rm gauge})+
({\rm fermions})\Bigr\rb
\lab{L1}
\ee
and similarly for $L_2$ (with different choice of the
normalization factors in front of each of the terms).  
Varying w.r.t. $\varphi^{i}$ and assuming $\Phi\neq 0$, we get:
\be
 L_{1}=sM^{4} = {\rm const} \; ,
\lab{varphi}
\ee
where  $s=\pm 1$ and $M$ has dimension of mass. The appearance of a nonzero integration
constant $sM^{4}$ {\em spontaneously breaks the scale invariance} \ct{G7}.

When including terms quadratic in the scalar curvature $R(\om,e)$, these types of models
can be applied not only for the late time universe, but also for
the early inflationary epoch. As it has been demonstrated in
Ref.\ct{G13}, a smooth transition between these epochs is
possible in these models. Also, these type of models provide the possibility of a
non-singular ``emergent'' type cosmology, where the existence and stability of 
singularity free universe imposes an upper bound on the cosmological constant today; 
for a review, see Ref.\ct{GL}.

\section{Extended objects}

Extended objects' actions can be formulated using a modified measure
analogous to \rf{Phi}. For simplicity we review here only the string case,
where on the 2-dimensional world-sheet we introduce:
\be
\Phi (\vp) =  \frac{1}{2}\varepsilon^{ab}  \varepsilon_{ij}
\partial_{a} \varphi^{i} \partial_{b} \varphi^{j} \; .
\lab{2-Phi}
\ee
In Ref.\ct{mstring} we have proposed the following modified-measure string action:
\be
S_{\rm mstring} =  - \h \int d^2\s \Phi(\vp) \Bigl\lb\g^{ab} \pa_a X^\m \pa_b X^\n g_{\m\n}
- \frac{\vareps^{ab}}{\sqrt{-\g}} F_{ab}\Bigr\rb \; ,
\lab{m-string}
\ee
where $F_{ab} = \pa_a A_b - \pa_b A_a$ with $A_{a}(\sigma)$ being an auxiliary
abelian world-sheet gauge field. Its presence is crucial for consistency of
the modified-measure string dynamics. Note that adding this term to the
standard Polyakov-type string action ($\Phi (\vp) \to \sqrt{-\g}$ in \rf{m-string})
would make it a purely topological (total divergence) term
$\h \int d^2\s\,\vareps^{ab} F_{ab}$.

The action \rf{m-string} is Weyl-conformally invariant under conformal rescaling of 
the world-sheet metric $\g_{ab} \to \g^{\pr}_{ab} = J \g_{ab}$ combined with a
diffeomorphism on the $\vp^i$-target space $\vp^i \to \vp^{\pr\,i} = \phi^i (\vp)$
with a Jacobian $\det\Vert \pa\phi^i/\pa\vp^j\Vert = J$.


The equation of motion obtained from variation of \rf{m-string} w.r.t. $A_{a}$ is 
$\varepsilon^{ab}\partial_{a}(\frac{\Phi}{\sqrt{-\gamma}}) = 0$, which yields a 
{\em spontaneously induced string
tension} $T=\frac{\Phi}{\sqrt{-\gamma}}={\rm const}$. The string tension appears here
as an integration constant and does not have to be introduced from the beginning. 
Let us stress that the string theory action \rf{m-string} 
does not have any \textsl{ad hoc} fundamental scale parameters. 

Variation of \rf{m-string} w.r.t. the measure fields
$\vp^i$ yields a fundamental constraint of the theory:
\be
g^{ab} \pa_a X^\m \pa_b X^\n g_{\m\n} - \frac{\vareps^{ab}}{\sqrt{-\g}} F_{ab}
= M \quad ,\quad M = {\rm const} \; ,
\lab{L-M}
\ee
which allows for 
$F_{ab}$ to be expressed in terms of the
basic string variables. Consistency of the whole set of equations of motion
demands $M=0$, so that finally we obtain the same equations as in the
standard bosonic string theory, however, with a dynamically induced
``floating'' string tension $T=\Phi/\sqrt{-\g}$. 

The above modified-measure formalism can be applied to the Green-Schwarz
superstring. As shown in Ref.\ct{mstring} the world-sheet gauge field $A_a$ plays 
crucial  role for ensuring supersymmetry invariance of the 
modified-measure superstring theory.

\section{Supergravity with Dynamically Induced Cosmological Constant}

The ideas and concepts of two-measure gravitational theories \ct{G6,G7} may be combined
with those originating from the theory of string and branes with
dynamical generation of string/brane tension \ct{mstring} to consistently
incorporate supersymmetry in the two-measure modification of standard
Einstein gravity. Here for simplicity we will present the modified-measure
construction of $N=1$ supergravity in $D=4$. For a recent account of
modern supergravity theories and notations, see Ref.\ct{freedman-proeyen}.

The standard component-field action of $D=4$ (minimal) $N=1$ supergravity
reads:
\br
S_{\rm SG} = \frac{1}{2\k^2} \int d^4 x\, e
\Bigl\lb R(\om,e) - {\bar\psi}_\m \g^{\m\n\l} D_\n \psi_\l \Bigr\rb \; ,
\lab{SG-action} \\
e = \det\Vert e^a_\m \Vert \;\; ,\;\;
R(\om,e) = e^{a\m} e^{b\n} R_{ab\m\n}(\om) \; .
\lab{curv-scalar} \\
R_{ab\m\n}(\om) = \pa_\m \om_{\n ab} - \pa_\n \om_{\m ab}
+  \om_{\m a}^c \om_{\n cb} - \om_{\n a}^c \om_{\m cb} \; .
\lab{curvature} \\
D_\n \psi_\l = \pa_\n \psi_\l + \frac{1}{4}\om_{\n ab}\g^{ab}\psi_\l \;\; ,\;\;
\g^{\m\n\l} = e^\m_a e^\n_b e^\l_c \g^{abc}
\lab{D-covariant}
\er 
where all objects belong to the first-order ``vierbein'' (frame-bundle) formalism,
\textsl{i.e.}, the vierbeins $e^a_\m$ (describing the graviton) and the 
spin-connection $\om_{\m ab}$ ($SO(1,3)$ gauge field acting on the gravitino
$\psi_\m$) are \textsl{a priori} independent fields (their relation arises
subsequently on-shell); $\g^{ab} \equiv \h \(\g^a \g^b - \g^b \g^a\)$
\textsl{etc.} with $\g^a$ denoting the ordinary Dirac gamma-matrices.
The invariance of the  action \rf{SG-action} under local supersymmetry 
transformations: 
\be
\d_\eps e^a_\m = \h {\bar\vareps}\g^a \psi_\m \;\; ,\;\;
\d_\eps \psi_\m = D_\m \vareps
\lab{local-susy}
\ee
follows from the invariance of the pertinent Lagrangian density up to a
total derivative: 
\be
\d_\eps \Bigl( e \bigl\lb R(\om,e) 
- {\bar\psi}_\m \g^{\m\n\l} D_\n \psi_\l \bigr\rb\Bigr) = 
\pa_\m \bigl\lb e\bigl({\bar\vareps}\z^\m\bigr)\bigr\rb \; ,
\lab{local-susy-L}
\ee
where $\z^\m$ functionally depends on the gravitino field $\psi_\m$.

We now propose a modification of \rf{SG-action} by replacing the standard measure
density $e = \sqrt{-g}$ by the alternative measure density $\P(\vp)$ \rf{Phi}:
\be
S_{\rm mSG} = \frac{1}{2\k^2} \int d^4 x\, \P(\vp)\,
\Bigl\lb R(\om,e) - {\bar\psi}_\m \g^{\m\n\l} D_\n \psi_\l 
+ \frac{\vareps^{\m\n\k\l}}{3!\,e}\, \pa_\m H_{\n\k\l} \Bigr\rb \; ,
\lab{mSG-action}
\ee
where a new term containing the field-strength of a 3-index
antisymmetric tensor gauge field $H_{\n\k\l}$ has been added. Note that its
inclusion in the standard supergravity action \rf{SG-action} would yield a
purely topological (total divergence) term like in the case of
modified-measure (super)string \ct{mstring} (cf. Eq.\rf{m-string}).

The equations of motion w.r.t. $H_{\n\k\l}$ and the ``measure'' scalars
$\vp^i$ read:
\br
\pa_\m \Bigl(\frac{\Phi(\vp)}{e}\Bigr) = 0 \;\; \to \;\;
\frac{\Phi(\vp)}{e} \equiv \chi = {\rm const} \; ,
\lab{chi-const} \\
R(\om,e) - {\bar\psi}_\m \g^{\m\n\l} D_\n \psi_\l
+ \frac{\vareps^{\m\n\k\l}}{3!\, e}\, \pa_\m H_{\n\k\l} = 2 M \; ,
\lab{L-const}
\er
where $M$ is an arbitrary integration constant.

Now it is straightforward to check that the modified-measure supergravity
action \rf{mSG-action} is invariant under local supersymmetry transformations
\rf{local-susy-L} supplemented by the transformation laws for $H_{\m\n\l}$
and $\Phi(\vp)$:
\be
\d_\eps H_{\m\n\l} = - e\,\vareps_{\m\n\l\k}\bigl({\bar\vareps}\z^\k\bigr)
\quad, \quad \d_\eps \Phi(\vp) = \frac{\Phi(\vp)}{e}\,\d_\eps e \; ,
\lab{local-susy-H-Phi}
\ee
which algebraically close on-shell, \textsl{i.e.}, when Eq.\rf{chi-const} is
imposed.

The role of $H_{\n\k\l}$ in the modified-measure action 
\rf{mSG-action} is to absorb, under local supersymmetry transformation,
the total derivative term coming from \rf{local-susy-L}, so as to insure
local supersymmetry invariance of \rf{mSG-action} -- this is a
generalization of the formalism used in Ref.\ct{mstring} to write down a
modified-measure extension of the standard Green-Schwarz world-sheet action
of space-time supersymmetric strings. Similar approach has also been
employed in Refs.\ct{nishino-rajpoot-1,nishino-rajpoot-2}. 

Let us particularly stress that the appearance of the integration constant $M$ 
in \rf{L-const} signifies a {\em spontaneous (dynamical) breaking of supersymmetry} 
and, simultaneously, it represents a {\em dynamically generated cosmological constant} in the
pertinent gravitational equations of motion. Indeed, varying \rf{mSG-action}
w.r.t. $e^a_\m$:
\br
e^{b\n}R^a_{b\m\n} - \h {\bar\psi}_\m \g^{a\n\l} D_\n \psi_\l
+ \h {\bar\psi}_\n \g^{a\n\l} D_\m \psi_\l
+ \h {\bar\psi}_\l \g^{a\n\l} D_\n \psi_\m
\nonu \\
+ \frac{e^a_\m}{2}\,\frac{\vareps^{\r\n\k\l}}{3!\, e}\, \pa_\r H_{\n\k\l} = 0
\lab{grav-eqs}
\er
and using Eq.\rf{L-const} to replace the last $H$-term on the l.h.s. of \rf{grav-eqs}
we obtain the vierbein analogues of the Einstein equations including a
dynamically generated {\em floating} cosmological constant term $e^a_\m M$:
\br
e^{b\n}R^a_{b\m\n} -\h e^a_\m R(\om,e) +  e^a_\m M = \k^2 T^a_\m \; ,
\nonu \\
\k^2 T^a_\m \equiv \h {\bar\psi}_\m \g^{a\n\l} D_\n \psi_\l
-\h e^a_\m {\bar\psi}_\r \g^{\r\n\l} D_\n \psi_\l
- \h {\bar\psi}_\n \g^{a\n\l} D_\m \psi_\l
- \h {\bar\psi}_\l \g^{a\n\l} D_\n \psi_\m \; .
\nonu \\
{}
\lab{einstein-eqs}
\er


\section*{Acknowledgments}
We gratefully acknowledge support of our collaboration through the academic exchange 
agreement between the Ben-Gurion University and the Bulgarian Academy of Sciences.
S.P. has received partial support from COST action MP-1210.


\end{document}